\documentclass[prl,preprint,amsmath]{revtex4}
\input {epsf.sty}
\usepackage{epsfig}
\usepackage{graphicx}
\usepackage{graphics}
\usepackage{amsmath,amsthm,amssymb,latexsym,amsfonts,times}
\usepackage{bm}
\usepackage{epstopdf} 

\begin{document}

\title{Microfluidic and Nanofluidic Cavities for Quantum Fluids Experiments}

\author{A. Duh$^1$, A. Suhel$^1$, B.D. Hauer$^1$, R. Saeedi$^1$, P.H. Kim$^1$, T.S. Biswas$^1$ and J.P. Davis$^{1,2}$}
\affiliation{1. Department of Physics, University of Alberta, Edmonton, Alberta, Canada T6G 2G7}
\affiliation{2. Canadian Institute for Advanced Research: Nanoelectronics Program, Toronto, Ontario, Canada M5G 1Z8}

\begin{abstract}

The union of quantum fluids research with nanoscience is rich with opportunities for new physics.  The relevant length scales in quantum fluids, $^3$He in particular, are comparable to those possible using microfluidic and nanofluidic devices.  In this article, we will briefly review how the physics of quantum fluids depends strongly on confinement on the microscale and nanoscale.  Then we present devices fabricated specifically for quantum fluids research, with cavity sizes ranging from 30 nm to 11 $\mu$m deep, and the characterization of these devices for low temperature quantum fluids experiments.

\end{abstract}

\maketitle

\section{Helium-4}

There are finite size effects near $T_c$ in nanoscale confined $^4$He, despite the fact that the relevant length scales in superfluid $^4$He are typically much smaller than possible through nanofabrication.  The zero temperature superfluid correlation length, $\xi(T=0)$, for $^4$He is approximately 0.1 nm \cite{Suk01}.  At $T=0$ the correlation length is essentially the same as the interatomic spacing between the $^4$He atoms, but it is temperature dependent and diverges at the superfluid transition temperature \cite{Suk01}, $\xi = \xi_0(1 - T/T_{c})^{-0.672}$.  It is this scaling of the correlation length with temperature that results in phenomena that show finite size effects in nanoscale geometries near $T_c$ \cite{Hos05}.  In particular, there are contributions to the heat capacity above $T_c$ that originate from the order parameter fluctuations.  As the temperature approaches $T_c$ from above, the correlation length increases until it diverges at $T_c$.  This has been explored in detail using heat capacity measurements by the Gasparini group at the University at Buffalo \cite{Gas08}.  Their experiments show that heat capacity is severely modified near $T_c$ in nanofabricated cavities, reducing the peak and broadening the transition \cite{Gas08}.  

To perform these experiments the Gasparini group fabricated microscale and nanoscale cavities using silicon wafers typically 5 cm in diameter \cite{Rhe90,Per10}.  They thermally oxidize these wafers to form a thin layer of silicon dioxide.  The thickness of this oxide layer determines the depth of the cavity and therefore they have fine control over their cavity sizes.

\section{Helium-3}

In order of decreasing length scale, some relevant length scales in liquid $^3$He are: the textural bending length of the $l$-vector in the $A$-phase ($\approx 10~\mu$m) \cite{Leg75}, the attenuation length of collisionless shear sound in the normal state (although temperature and frequency dependent it is expected to be on the order of microns for MHz ultrasound \cite{Hal90}), and the superfluid coherence length ($\xi_0 =$ 80-20 nm from 0-34 bar \cite{Whe75}).  All of these lengths are on scales that can be experimentally modified using nanofabricated cavities.  We are motivated in our device fabrication by the desire to control the latter two of these length scales.  

In Landau's work on Fermi liquid theory, he predicted that there would be a propagating transverse sound mode in the normal state of $^3$He \cite{Lan56,Lan57}.  This arises because of the collective dynamics of the strongly correlated Fermi liquid.  Transverse sound in the normal state is a nonhydrodynamic collisionless sound mode ($\omega\tau \geq 1$) and is expected to propagate at low temperatures and high frequencies with a restoring force from the Landau molecular field \cite{Moo93}.  Transverse sound has been observed as a propagating mode in the superfluid state through coupling to the order parameter collective modes \cite{Kal93,Lee99,Dav08a,Dav08d}, which is, in part, a consequence of transverse sound in the normal state.  But despite transverse acoustic measurements that have measured attenuation as high as 1000 cm$^{-1}$ \cite{Dav08f} (corresponding to an attenuation length of 10 $\mu$m) there has yet been no direct measurement of this propagating mode.  This is likely due to the extremely high attenuation of the transverse mode and the fact that the acoustic cavities formed have only been as small as 31 $\mu$m \cite{Dav08b}.  The much smaller acoustic path lengths in the devices we present below should be able to directly reveal transverse sound in the normal state.

In addition, there is a recent theoretical prediction based on confinement of $^3$He on the scale of the superfluid coherence length of a new superfluid phase that breaks translational symmetry: a crystalline superfluid \cite{Vor07}.  It is predicted that this phase will occur when the liquid is confined to approximately $10 \times $ the coherence length.  It forms because of a change in sign of the $z$-component of the $B$-phase order parameter when a Cooper-pair is reflected from a surface, resulting in a pair-breaking mechanism \cite{Vor03}.  Formation of domain walls \cite{Vor05} between two $B$-phase superfluids with alternating sign of the $z$-component of the order parameter, which are otherwise degenerate, eliminates this pair-breaking mechanism.  But this is at the expense of pair-breaking upon crossing the domain wall.  As a result, there  will form a superfluid with stripes or checkerboard domains, which is expected to be stable, but which will only occur within a narrow range of cavity sizes \cite{Vor07}.

Recently, experiments have begun to pursue the detection of this state.   Sensitive SQUID NMR is being performed in the Saunders Group at Royal Holloway \cite{Ben10,Lev10} along with torsional resonator measurements in the Parpia group at Cornell \cite{Dim10a} using silicon and glass microfabricated cavities \cite{Dim10b}.  In addition, the micro-electro-mechanical devices (MEMS) of the Lee group at University of Florida \cite{Gon10} have gaps between the substrate and the shear MEMS device of 1.25 and 0.75 $\mu$m, which are within the predicted range of the striped phase.  These may be able to detect the striped phase through changes in the response of the MEMS as a result of the phase transition.  We intend to search for this state using an acoustic probe, as described in the next section.

Finally, nanofabricated devices may be able to directly confront some of the observations made in thin films of $^3$He \cite{Dav88,Sch98,Xu90} without the complication of nonuniform film thickness.

\section{Proposed Experimental Arrangement}

We propose an experimental arrangement using nanofabricated devices that will be able to explore finite size effects in $^4$He near $T_c$, as well as transverse sound in the normal state of $^3$He, and search for the crystalline superfluid state of $^3$He.  The central feature of this arrangement is a nanofabricated glass microfluidic or nanofluidic device of a 2D geometry.  Such a device can be probed by exciting a standing acoustic wave in the out-of-plane direction of the device using an ultrasonic transducer bonded to one side of the device.  When the ultrasound velocity or attenuation is varied, by tuning either the pressure or temperature, a change in the standing wave is detected as a change in the acoustic impedance of the transducer.  This has been used successfully in both longitudinal \cite{Dav08c} and transverse sound measurements of $^3$He in devices down to $31~\mu$m \cite{Lee99,Dav08a,Dav08d,Dav08f,Dav08b}, which were not limited by signal-to-noise but instead by the ability to generate smaller cavities.  

Ultrasound is well suited to explore the above phenomena, which is clear for measurements of transverse sound in the normal state of $^3$He.  In addition, it has been shown that Andreev bound states near a wall can be probed using transverse acoustic impedance \cite{Dav08f,Wad08} and that the domain walls formed in the crystalline superfluid state are essentially Andreev bound states \cite{Vor05}.  Therefore the crystalline superfluid state will have a signature in ultrasound attenuation from the increase attenuation of the domain walls.  For experiments in $^4$He, the signatures of the lambda transition are also revealed in measurements of the velocity and attenuation of ultrasound \cite{Bar68,Lam79} and broadening of these features in the nanofabricated devices can be compared with measurements of bulk $T_c$.

A feature of the proposed arrangement is that the nanofluidic device does not need to be independently sealed, but instead can be submersed directly into the liquid helium.  This avoids any concern over deformation of the cavity walls (and therefore change in the cavity size) under pressurization.  With clear signatures of the proposed quantum fluids phenomena, as well as an established experimental acoustic impedance technique, the challenge in the proposed experimental arrangement is the fabrication of appropriately sized glass microfluidic or nanofluidic devices, which also meet the needs of low temperature experiments.  We have now successfully completed fabrication of such devices, as described below.  

\section{Glass Microfluidic and Nanofluidic Cavities}
\subsection{Fabrication}

The basic design of our micro/nanofluidic cavities is a basin (a cylindrical, quasi-2D cavity) of 3, 5 or 7 mm in diameter, with two channels leading from the edge of the device into the basin.  The channel widths are varied depending on the targeted etch depth (thinner cavities have wider channels to ensure proper fluid access to the basin).  The channels eliminate any bulk fluid volumes near the micro/nanofluidic cavities, which is vital to our proposed acoustic measurements, but present a fabrication difficulty.  Typically glass microfluidic devices have access holes drilled in the top capping glass piece.  This allows the channels to terminate before the edge of the device and therefore they can be fabricated on a wafer scale and diced after bonding \cite{Dim10b}.  With our design, dicing after bonding would result in glass dust from dicing clogging the channels and prohibiting fluid access.  Instead, our devices are diced and etched individually before sealing.

\begin{figure*}[t]
%%%%%%%%%%%%%%%%%   F I G U R E  1   %%%%%%%%%%%%%%%%%%
 \centerline{\includegraphics[width=1\textwidth]{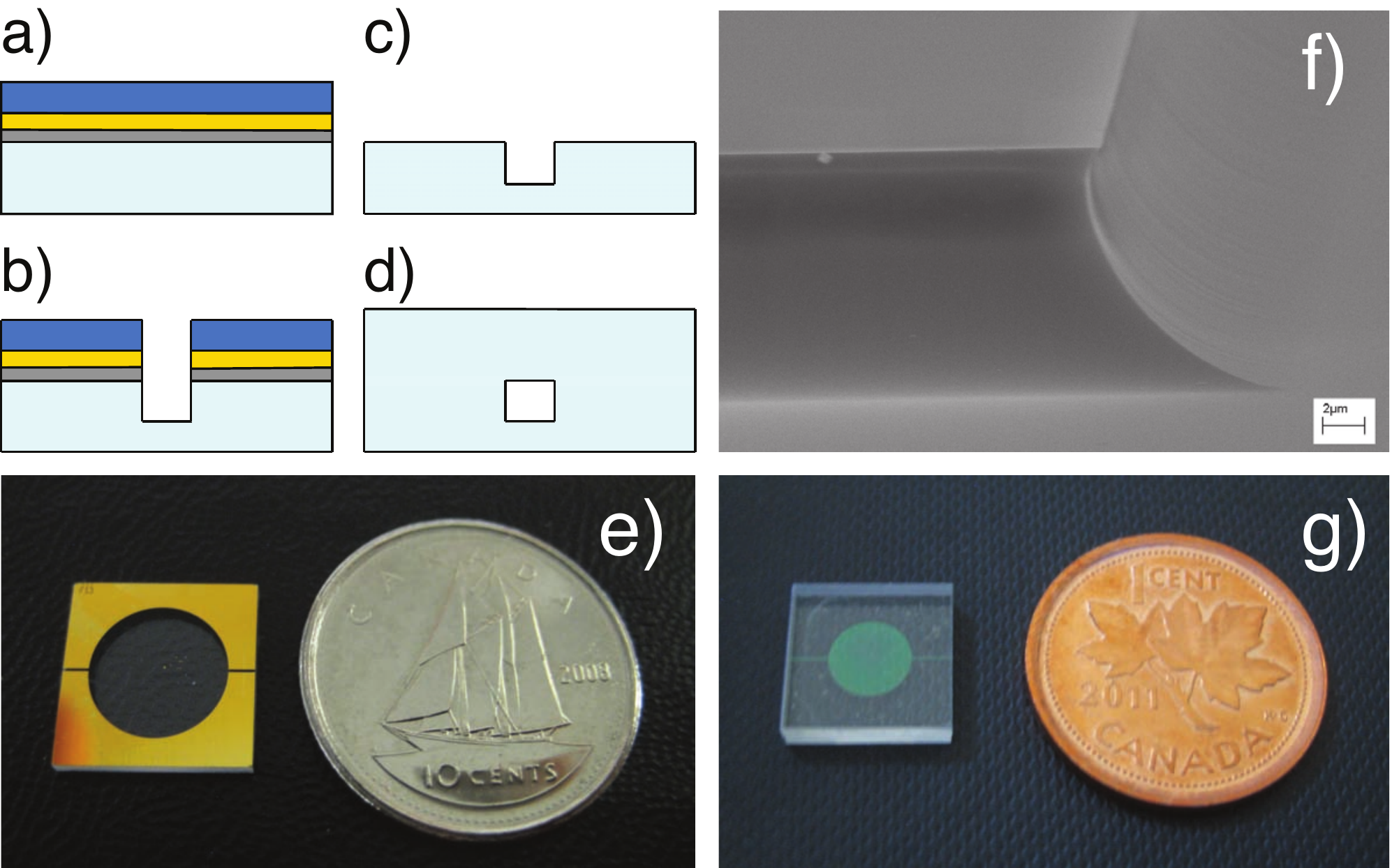}}
%%%%%%%%%%%%%%%%%%%%%%%%%%%%%%%%%%%%%%%%%%%%%
\caption{{\label{fig1}}  Process flow and device characterization.  Panels (a) through (d) show a simplified nanofabrication procedure, involving: applying metal masking layers and photoresist (a); photoresist developing, metal etching and silicon dioxide etching (b); removal of masking layers (c); and glass bonding and annealing (d).   e) Optical photograph of a 7 mm diameter basin device after step (b).  f) SEM image of a 9 $\mu$m deep device, at the location that the channel meets the basin.  This shows the precision of the optical lithography.  g) Optical photograph of a fully processed 5 mm diameter basin, 300 nm deep device.  The green color is a result of thin film interference.  The uniformity of the color is a testament to the cavity depth uniformity.}
\end{figure*}

The devices presented here were fabricated using optical lithography at the University of Alberta NanoFab.  The starting material was 4 inch $\times$ 4 inch $\times$ 0.043 inch square borofloat (81\% SiO$_2$,13\% B$_2$O$_2$, 4\% Na$_2$O/K$_2$O, 2\% Al$_2$O$_3$) glass wafers.  These were piranha cleaned prior to depositing metal masking layers.  We used 30 nm of Cr as an adhesion layer and 180 nm of Au as the masking layer.  At this point the photoresist (HPR504) is spun onto the wafer for 10 seconds at 500 rotations per minute (RPM) and then 40 seconds at 4000 RPM and  baked for 30 minutes, Fig.1a.  The resulting thickness of the photoresist is 1.2 $\mu$m as measured by an Alpha Step profilometer.  The photoresist is then exposed to 2.5 seconds of 365 nm light at 89.50 mW/cm$^2$ through the photomask.  The photomask was designed using L-edit software and fabricated on a Heidelberg DWL-200 pattern generator.  Exposure transfers the photomask pattern to the photoresist, which is then developed to remove the exposed pattern.  Next the Au and Cr layers are wet etched with chemically specific etchants, KI, and a mixture of ceric ammonium nitrate and nitric acid, respectively.  This exposes the borofloat in precise locations as dictated by the photomask. 

We then dice the borofloat wafer into individual devices, so that each can be etched to a different depth.  We also dice a blank borofloat wafer to match the individual devices, which will later be used as the top sides of the cavities.  The devices are then etched in a borofloat etch (50\% HF, 10\% nitric acid and 40\% water) for a specified time (at 12.5 nm/s), which determines the final cavity size.  Once this step is complete the photomask pattern has been transferred to the borofloat, Fig. 1b.  A photograph of a device at this step is shown in Fig. 1e.  At this stage we perform extensive analysis of the depth and roughness of the cavity formed in the borofloat, which we describe in the next section.  Afterwards, the photoresist, Au and Cr are stripped from the device, Fig. 1c.  An SEM image of a 9 $\mu$m deep cavity at this stage is shown in Fig. 1f.  The cavity edges are very clearly defined because of the metal masking layer, and the walls have a characteristic rounding from the isotropic borofloat etch.  Thinner cavities show less rounding.

Finally, the etched devices are piranha cleaned, bonded to the blank borofloat pieces and annealed at $600 \,^{\circ}\mathrm{C}$ for 120 minutes (the temperature is raised and lowered at $10 \,^{\circ}\mathrm{C}$/minute).  It is essential that the devices be extremely clean before bonding, to ensure a good bond between the two pieces of borofloat.  After annealing the two pieces of borofloat become one, Fig. 1d, and it is impossible to separate them at the bond.  As can be readily seen in the completed device in Fig. 1g, the cavities show strong thin film interference from the cavity size.  The uniformity of the reflected color demonstrates that the cavities have a uniform depth across the diameter of the basin.

\begin{figure*}[t]
%%%%%%%%%%%%%%%%%   F I G U R E  2   %%%%%%%%%%%%%%%%%%
 \centerline{\includegraphics[width=1\textwidth]{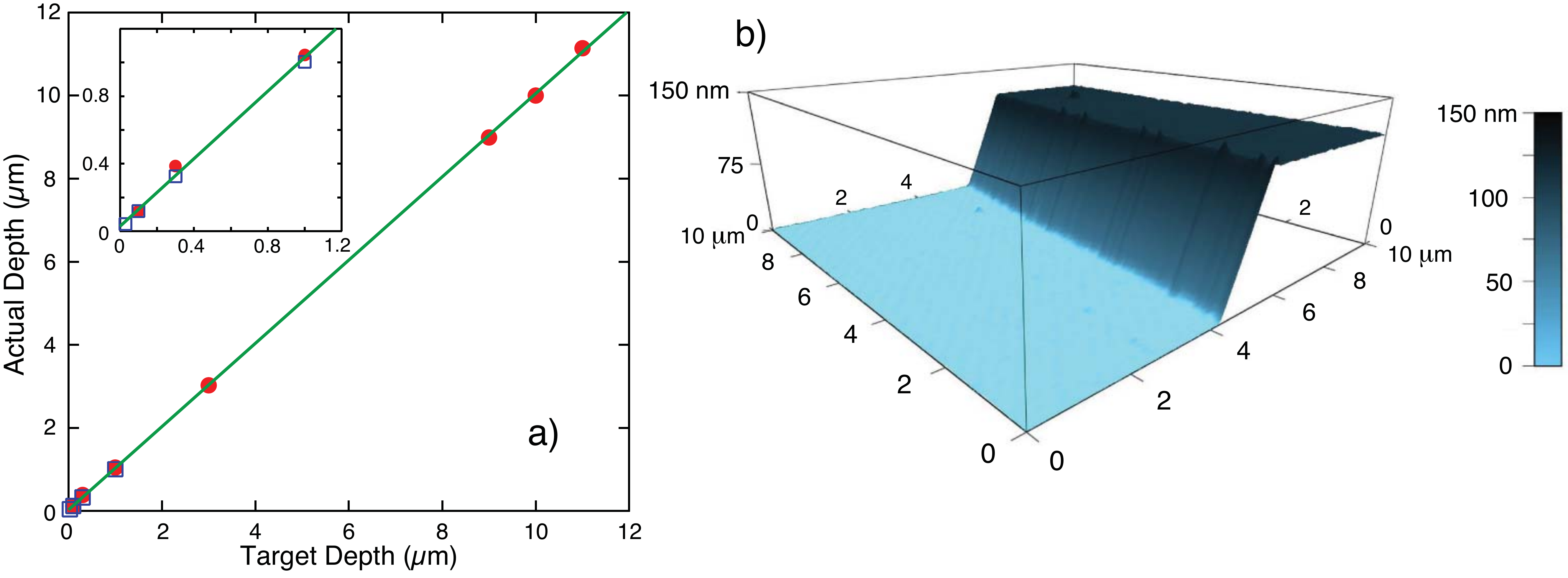}}
%%%%%%%%%%%%%%%%%%%%%%%%%%%%%%%%%%%%%%%%%%%%%
\caption{{\label{fig2}} a) Actual etch depth versus target etch depth as measure by profilometer (red circles) and AFM (blue squares).  A fit to the combined data (green line) has a slope of 1.002, verifying that our target depth to actual depth is precise.  b) An image of the AFM surface topography of a 100 nm deep device.  }
\end{figure*}

\subsection{Cavity Characterization}

For the purposes of quantum fluids experiments it is important to know the thickness of the cavity that will define the superfluid.  Therefore we have performed extensive characterization of our cavity depths prior to the bonding and annealing step.  In Fig. 1f, we show an SEM image of a typical device, which reveals the well defined cavity walls.  The depth is varying at the wall, which will not adversely effect an experiment that is sensitive to the entire volume, unless the texture near the wall contributes to the experimental signal.  The cavity depths have been measured by an Alpha Step profilometer, but the accuracy of the profilometer degrades for the smallest devices, and the 30 nm cavity is not measurable with the profilometer.  To solve this problem, we also measure the cavity depth using a calibrated AFM (Asylum MFP-3D) on devices 1 $\mu$m deep and smaller.  The combined profilometer and AFM data, Fig. 2a, confirms that we are able to accurately fabricate devices anywhere from 30 nm to 11 $\mu$m deep.  Thirty nanometers is approaching the size scale in which one might expect phenomenon related to two dimensional superfluidity in liquid $^3$He, like a Kosterlitz-Thouless transition \cite{Kos73} or the Stein-Cross transition to islands of singularities in the $l$-vector \cite{Ste79}.  In Fig. 2b we show an AFM surface topography of a 100 nm deep cavity.  The AFM does not reveal the rounded wall shape that is visible in the SEM.  This is due to the thinner cavity, and the inability of AFM to measure the steep slope of the wall due to the profile of the cantilever tip.  In the AFM topography there is a clear, sharp transition from the unetched borofloat above and the etched basin below.

\begin{figure*}[t]
%%%%%%%%%%%%%%%%%   F I G U R E  3   %%%%%%%%%%%%%%%%%%
\centerline{\includegraphics[width=1\textwidth]{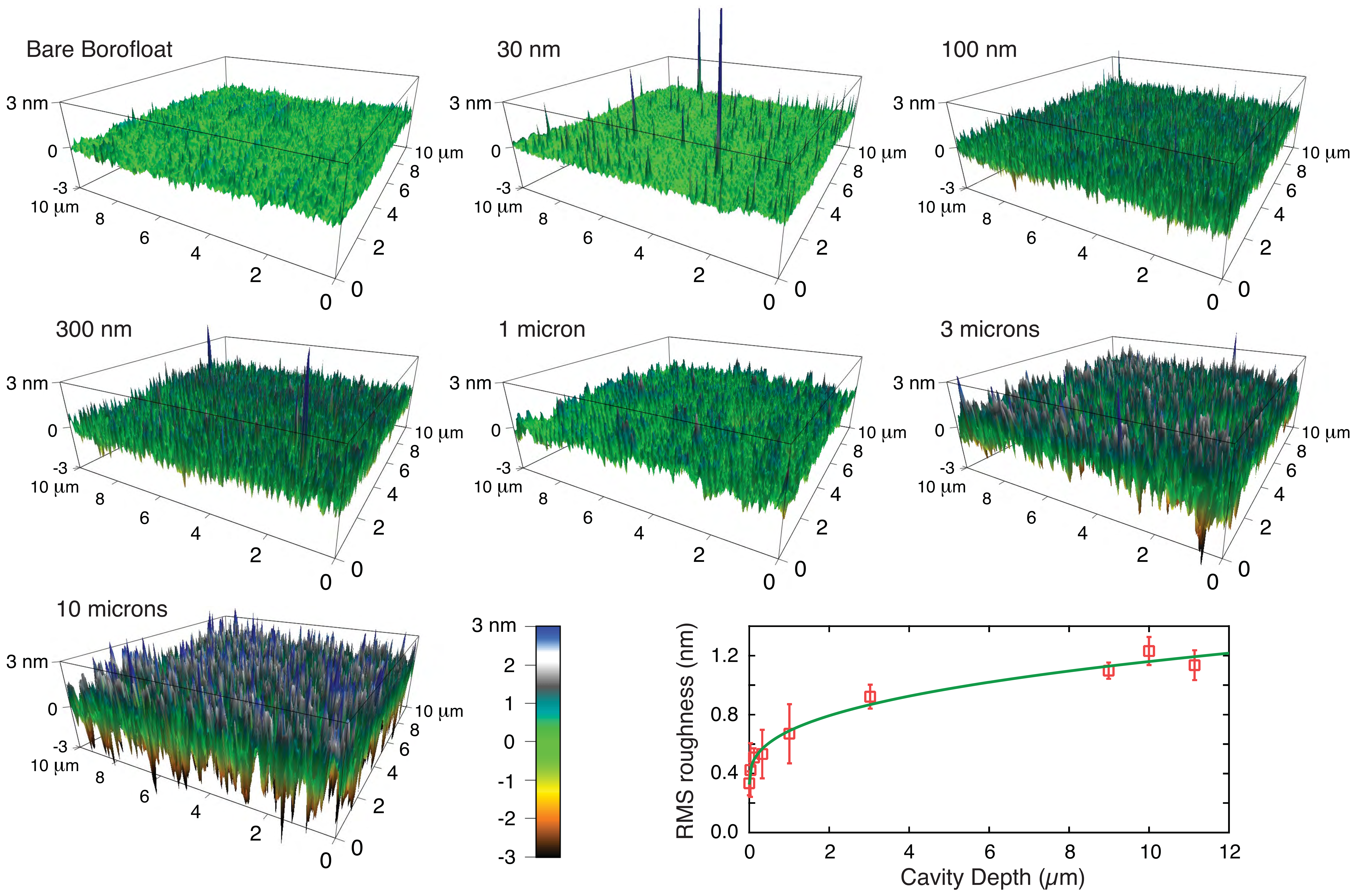}}
%%%%%%%%%%%%%%%%%%%%%%%%%%%%%%%%%%%%%%%%%%%%%
\caption{{\label{fig3}} AFM surface topography measurements inside the etched basins labeled with their etch depth and a plot of the RMS surface roughness versus etch depth.  The roughness increases with increasing etch depth, yet all devices have an RMS roughness below $\approx$ 1.2 nm.  The green curve is a guide to the eye. }
\end{figure*}

After bonding and annealing it is difficult to optically measure the cavity depth and uniformity in our transparent glass devices.  We have attempted such measurements using a commercial Zygo Optical Profilometer, as well as a custom built scanning optical interferometer, but without success.  There is insufficient reflection from the cavity surfaces for such measurements.  That being said, we can place bounds on the uniformity of our devices using thin film interference, as seen in Fig. 1g.  A wavelength change of 5 nm in the blue-green color range is obvious to the eye.  Therefore we can set an approximate upper bound of 2.5 nm across the radius of the basin, which is a 2.5\% variation in a 100 nm thick device and less than 1\% in cavities of 300 nm and larger.

Another parameter in the study of quantum fluids is the surface roughness \cite{Sha11}.  In particular, specular and diffuse scattering surfaces suppress the order parameter differently \cite{Leg75,Amb74} and therefore in the theory of Vorontsov and Sauls the surface scattering alters the phase diagram of the crystalline superfluid phase in nanoscale cavities \cite{Vor03}.  We have characterized the surface roughness of our etched basins and find a clear trend with etch depth, Fig. 3.  The 30 nm cavity has a roughness that is very similar to the unadulterated borofloat wafer and the thickest devices approach 1.2 nm RMS roughness.  This is comparable to the 1.01 nm standard deviation in the surface height for the Hoya SD-2 glass of Ref. \cite{Dim10b}.  

Currently, the theory of Vorontsov and Sauls \cite{Vor07,Vor03} is only for fully specular or fully diffuse scattering and cannot account for intermediate roughness as in our devices \cite{Sau11}.  It is therefore not possible at this time to say quantitatively where along the spectrum of specular to diffuse scattering our devices lie.  But these surfaces are sufficiently smooth that with a preplating of 2-3 atomic layers of $^4$He we will achieve fully specular scattering \cite{Fre88}.  We have also attempted other recipes for the borofloat etch than the one listed above, including buffered oxide etch (BOE) (10:1 mixture of HF and H$_5$F$_2$N) and reactive ion etching (RIE), all of which resulted in much rougher surfaces.

We have tested the cavities to ensure that the channels leading from the edge of the wafer are in fact unblocked after bonding and annealing.  Upon application of dye to the channel entry the basin fills with dye, which is clearly visible by eye.  This is true for all cavity depths, including the thinnest cavity devices.

Finally, we have checked that the devices withstand temperature cycling.  Plunging devices into liquid nitrogen and subsequent warming to room temperature, even rapidly, has no adverse effect on the devices, nor does multiple temperature cyclings.  This is true whether the channels are open, or intentionally blocked.  This is a consequence of the strong bond formed between the two glass pieces upon bonding and annealing, in essence forming a single piece of glass.  Therefore we are confident that these devices will withstand the ultra-low temperatures of our proposed quantum fluid experiments.

\section{Conclusion}

We have designed and fabricated glass (borofloat) microfluidic and nanofluidic devices for quantum fluids research with no bulk fluid in the vicinity of the micro/nanofluidic cavity.  We can accurately etch the fluid cavities to any depth between 30 nm and 11 $\mu$m.  The devices are robust at low temperatures and will serve as the platform for acoustic impedance experiments of quantum fluids in precisely defined confined geometries.

\section{Acknowledgments}

This work was supported by the University of Alberta, Faculty of Science; the Natural Sciences and Engineering Research Council, Canada; and the Canada Foundation for Innovation.  We would like to thank  W.P. Halperin, J.R. Beamish, M.R. Freeman, H. Choi, J. Pollanen,  L. Li and W.J. Gannon for stimulating discussions.  We are grateful to the technical support of Greg Popowich and Don Mullin, and the staff of the University of Alberta NanoFab for their assistance in device fabrication.

\end{document}